\documentclass[aps,prd,showpacs,nofootinbib,tightenlines]{revtex4}
\usepackage{amsmath}
\usepackage{amssymb}
\usepackage{epsfig}
\usepackage{graphicx}
\begin{document}

\title{Quasi-two-body decays $B\to K\rho\to K\pi\pi$ in perturbative QCD approach}
\author{Wen-Fei Wang$^{1,2}$}\email{wfwang@sxu.edu.cn}
\author{Hsiang-nan Li$^1$}\email{hnli@phys.sinica.edu.tw}
\affiliation{$^1$Institute of Physics, Academia Sinica, Taipei, Taiwan 115, Republic of China}
\affiliation{$^2$Institute of Theoretical Physics, Shanxi University, Taiyuan, Shanxi 030006, China}

\date{\today}

\begin{abstract}
  We analyze the quasi-two-body decays $B\to K\rho\to K\pi\pi$ in the perturbative QCD (PQCD)
  approach, in which final-state interactions between the pions in the resonant regions associated
  with the $P$-wave states $\rho(770)$ and $\rho^\prime(1450)$ are factorized into two-pion distribution
  amplitudes. Adopting experimental inputs for the time-like pion form factors involved in two-pion
  distribution amplitudes, we calculate branching ratios and direct $CP$ asymmetries of the
  $B\to K\rho(770),K\rho^\prime(1450)\to K\pi\pi$ modes. It is shown that agreement of theoretical
  results with data can be achieved, through which Gegenbauer moments of the $P$-wave two-pion
  distribution amplitudes are determined. The consistency between the three-body and two-body
  analyses of the $B\to K\rho(770)\to K\pi\pi$ decays supports the PQCD factorization framework for exclusive 
  hadronic $B$ meson decays.
\end{abstract}

\pacs{13.20.He, 13.25.Hw, 13.30.Eg}
\maketitle

\section{INTRODUCTION}

Strong dynamics contained in three-body hadronic $B$ meson decays is much more
complicated than in two-body cases, because of entangled nonresonant and
resonant contributions, and significant final-state interactions~\cite{1512-09284}.
Nonresonant contributions may not be negligible in these decays,
as indicated by the observations made in Refs~\cite{prd80-112001,
prd79-072006,prd78-052005,prd75-012006,prl96-251803,prl77-4503}.
Quasi-two-body channels through intermediate scalar, vector
and tensor resonances, which produce hadron pairs with final-state
interactions, usually dominate total branching fractions. 
An amplitude for a three-body hadronic $B$ meson decay, as
a coherent sum of nonresonant and resonant contributions,
leads to nonuniform distributions of events described by differential
branching fractions~\cite{prd80-112001,prd79-072006,
prd78-052005,prd75-012006,prl96-251803,prl77-4503,prd79-072004,prd71-092003,
prd65-092005,prd78-012004,prl99-221801,prd74-032003,prd72-072003,prd72-052002,
prd70-092001} and of direct $CP$ asymmetries~\cite{prd90-112004,prl112-011801,
prl111-101801,prl99-161802} in a Dalitz plot~\cite{dalitz-plot}.
Dalitz-plot analyses of abundant three-body hadronic $B$
meson decays from different collaborations (BaBar~\cite{prd80-112001,prd79-072006,
prd78-052005,prd78-012004,prl99-221801,prd74-032003,prd72-072003,prd72-052002,
prd70-092001}, Belle~\cite{prd75-012006,prl96-251803,prd79-072004,prd71-092003,
prd65-092005} and LHCb~\cite{prd90-112004,prl112-011801,prl111-101801}) have revealed
valuable information on involved strong and weak dynamics.

On the theoretical side, substantial progress on three-body hadronic $B$ meson decays 
by means of symmetry principles and factorization theorems has been made, 
although rigorous justification of these approaches is not yet available. Isospin, 
U-spin and flavor SU(3) symmetries were adopted 
in~\cite{plb564-90,prd72-075013,prd72-094031,prd84-056002,plb727-136,plb726-337,
prd89-074043,plb728-579,ijmpa29-1450011,prd91-014029}, and the role of the $CPT$ 
invariance in three-body $B$ meson decays was discussed in 
Refs~\cite{prd89-094013,prd92-054010}. The QCD factorization~\cite{QCDF-I,QCDF-PV}
has been widely applied to studies of three-body charmless hadronic $B$ meson
decays~\cite{plb622-207,prd74-114009,prd79-094005,prd81-094033,appb42-2013,
prd66-054015,prd72-094003,prd76-094006,prd88-114014,prd89-074025,prd89-094007,
epjc75-536,prd87-076007}, including, for instance, detailed investigation 
on factorization properties of the $B^+\to\pi^+\pi^+\pi^-$ mode in various 
regions of phase space~\cite{npb899-247}. The perturbative QCD (PQCD)
approach based on the $k_T$ factorization theorem~\cite{plb504-6+prd63-054008,prd63-074009}
has been employed in Refs.~\cite{plb561-258,prd70-054006,prd89-074031,prd91-094024,
arXiv1509-06117}, where strong dynamics between two final-state hadrons in
resonant regions are factorized into a new nonperturbative input, the two-hadron
distribution amplitudes. An advantage of the PQCD factorization approach is that 
both nonresonant and resonant contributions can be accommodated into this new input. 
A model that combines the heavy quark effective theory and the chiral Lagrangian
was proposed in Ref.~\cite{prd70-034033} to compute nonresonant decay amplitudes.
The $B$ meson transition to a meson pair has been analyzed in the heavy-mass and 
large-energy limits~\cite{Faller:2013dwa}, and in the light-cone sum
rules~\cite{Hambrock:2015aor} also in terms of two-meson distribution amplitudes. 
Nonresonant contributions to the above transition were evaluated in the heavy meson
chiral perturbation theory~\cite{hmchpt} 
in~Refs.~\cite{prd89-094007,prd76-094006,prd88-114014}. 

In this Letter we will focus on resonant contributions to three-body hadronic
$B$ meson decays in the PQCD approach, extending our previous work on
$S$-wave resonances to $P$-wave ones. We have determined the Gegenbauer moments
of the $S$-wave two-pion distribution amplitudes by fitting our formalism to the
$B^0_{(s)}\to J/\psi\pi^+\pi^-$ and $B_s\to\pi^+\pi^-\ell^+\ell^-$
data. Here we will consider the quasi-two-body decays $B\to K\rho\to K\pi\pi$, 
which receive contributions mainly from the $\rho(770)$ and $\rho^\prime(1450)$ 
intermediate states. These resonant contributions are parametrized into the 
time-like pion form factors involved in the two-pion distribution amplitudes, 
for which there exist experimental inputs from the $e^+e^-$ annihilation. It will 
be demonstrated that agreement of theoretical results with data can be achieved 
by choosing appropriate Gegenbauer moments of the $P$-wave two-pion distribution
amplitudes. On one hand, the consistency between the three-body and two-body
analyses of the quasi-two-body modes $B\to K\rho(770)\to K\pi\pi$ to be verified below
supports the PQCD factorization for exclusive hadronic $B$ meson decays. 
On the other hand, with both the $S$-wave and $P$-wave distribution amplitudes
being ready, we can proceed to predictions for branching ratios and direct $CP$ 
asymmetries of three-body hadronic $B$ meson decays in various localized regions
of two-pion phase space.

The rest of this Letter is organized as follows. The PQCD framework for three-body
hadronic $B$ meson decays is reviewed in Sec.~II, where the $P$-wave two-pion
distribution amplitudes up to twist 3 are parametrized. Numerical results for 
branching ratios and direct $CP$ asymmetries of the various 
$B\to K\rho\to K\pi\pi$ modes are presented and compared with those from the
two-body analysis in Sec.~III. The straightforward extension
of the present formalism to other $P$-wave resonant contributions is highlighted.
Section~IV contains the Conclusion. The factorization formulas for the relevant
three-body decay amplitudes are collected in the Appendix.

\section{FRAMEWORK}

In the rest frame of the $B$ meson, we write the $B$ meson momentum $p_B$
and the light spectator quark momentum $k_B$ as
\begin{eqnarray}
p_B=\frac{m_B}{\sqrt2}(1,1,0_{\rm T}),\quad
k_B=\left(0,\frac{m_B}{\sqrt2}x_B ,k_{B{\rm T}}\right),
\end{eqnarray}
in the light-cone coordinates,
with $m_B$ being the $B$ meson mass and $x_B$ the momentum fraction.
For the $B\to K\rho\to K\pi\pi$ decays, we define the resonant state
momentum $p$ (in the plus $z$ direction) and the associated spectator quark 
momentum $k$, and the kaon momentum $p_3$ (in the minus $z$ direction)
and the associated non-strange quark momentum $k_3$ as
\begin{eqnarray}\label{def-pp3}
p=\frac{m_B}{\sqrt2}(1,\eta,0_{\rm T}),\quad
k=\left(\frac{m_B}{\sqrt2}z,0,k_{\rm T}\right),\quad
p_3=\frac{m_B}{\sqrt2}(0,1-\eta,0_{\rm T}),\quad
k_3=\left(0,\frac{m_B}{\sqrt2}(1-\eta)x_3,k_{3{\rm T}}\right),
\end{eqnarray}
with the variable $\eta=w^2/m^2_B$, $w=\sqrt{p^2}$ being the invariant mass
of the resonant state, and the momentum fractions $z$ and $x_3$.
The momenta $p_1$ and $p_2$ for the two pions from the resonant state have
the components~\cite{prd89-074031}
\begin{eqnarray}
p^+_1=\zeta\frac{m_B}{\sqrt2},           \quad    p^-_1=(1-\zeta)\eta\frac{m_B}{\sqrt2},
\quad p^+_2=(1-\zeta)\frac{m_B}{\sqrt2}, \quad    p^-_2=\zeta\eta\frac{m_B}{\sqrt2},
\label{def-pp4}
\end{eqnarray}
in which the momentum fraction $\zeta$ of the first pion runs between 0 and 1.

In Ref.~\cite{prd91-094024} we have introduced the distribution amplitudes for the pion
pair~\cite{Mueller:1998fv,Diehl:1998dk,npb555-231}
\begin{eqnarray}
\phi_{v\nu}^I(z,\zeta,w^2)&=&\frac{1}{2\sqrt{2N_c}}\int \frac{dy^-}{2\pi}
e^{-izp^+ y^-}\langle\pi^+(p_1)\pi^-(p_2)|\bar\psi(y^-)\gamma_\nu T \psi(0)|0\rangle, \label{pa}\\
\phi_s^I(z,\zeta,w^2)&=&\frac{1}{2\sqrt{2N_c}}\frac{p^+}{w} \int \frac{dy^-}{2\pi}
e^{-izp^+ y^-}\langle\pi^+(p_1)\pi^-(p_2)|\bar\psi(y^-)T \psi(0)|0\rangle, \label{ps}\\
\phi_{t\nu}^I(z,\zeta,w^2)&=&\frac{1}{2\sqrt{2N_c}}\frac{p^+f_{2\pi}^\perp}{w^2}
\int \frac{dy^-}{2\pi} e^{-izp^+ y^-}\langle\pi^+(p_1)\pi^-(p_2)|\bar\psi(y^-)
i\sigma_{\mu\nu}n_-^\mu  T\psi(0)|0\rangle, \label{pt}
\end{eqnarray}
where $N_c$ is the number of colors, $n_-=(0,1,{\bf 0}_T)$ is a dimensionless vector, 
$T=\tau^3/2$ is chosen for the isovector $I=1$ state, $\psi$ represents the $u$-$d$ 
quark doublet, and $f_{2\pi}^\perp$ is a normalization constant.
For $I=1$, the $P$-wave is the leading partial wave, to which
$\phi_{v\nu=-}^{I=1}$ and $\phi_{t\nu=\perp}^{I=1}$ contribute at twist 2, and
$\phi_{v\nu=\perp}^{I=1}$, $\phi_s^{I=1}$, and $\phi_{t\nu=+}^{I=1}$ contribute at twist 3.
With $w^2$ being a variable, the above two-pion distribution amplitudes
contain both nonresonant and resonant contributions from the
pion pair.

The $P$-wave two-pion distribution amplitudes are organized into
\begin{eqnarray}
\phi_{\pi\pi}^{I=1}=\frac{1}{\sqrt{2N_c}}\left[{p\hspace{-1.5truemm}/}
  \phi_{v\nu=-}^{I=1}(z,\zeta,w^2)+w\phi_{s}^{I=1}(z,\zeta,w^2)
  +\frac{{p\hspace{-1.5truemm}/}_1{p\hspace{-1.5truemm}/}_2
  -{p\hspace{-1.5truemm}/}_2{p\hspace{-1.5truemm}/}_1}{w(2\zeta-1)}
  \phi_{t\nu=+}^{I=1}(z,\zeta,w^2)
  \right]\;,
\end{eqnarray}
whose components are parametrized as
\begin{eqnarray}
\phi_{v\nu=-}^{I=1}(z,\zeta,w^2)\equiv \phi^{0}(z,\zeta,w^2)
&=&\frac{3F_\pi(w^2)}{\sqrt{2N_c}}
z(1-z)\left[1+a_2^{0} C^{3/2}_2(1-2z)\right]P_1(2\zeta-1)\;,\label{def-DA0}\\
\phi_{s}^{I=1}(z,\zeta,w^2)\equiv \phi^{s}(z,\zeta,w^2)
&=&\frac{3F_s(w^2)}{2\sqrt{2N_c}}
(1-2z)\left[1+a_2^s\left(1-10z+10z^2\right) \right]P_1(2\zeta-1)\;,\\
\phi_{t\nu=+}^{I=1}(z,\zeta,w^2)\equiv \phi^{t}(z,\zeta,w^2)
&=&\frac{3F_t(w^2)}{2\sqrt{2N_c}}
(1-2z)^2\left[1+a_2^t C^{3/2}_2(1-2z)\right]P_1(2\zeta-1)\;,\label{def-DAs}
\end{eqnarray}
with the Gegenbauer polynomial $C^{3/2}_2(t)=3\left(5t^2-1\right)/2$ and
the Legendre polynomial $P_1(2\zeta-1)=2\zeta-1$. In principle, the time-like
form factors associated with the second Gegenbauer moments $a_2^{0,t,s}$ can differ
from $F_{\pi,s,t}(w^2)$ associated with the leading ones. Here we assume that they
are the same, which can then be factored out and serve as the normalization of
the two-pion distribution amplitudes. The moments $a_2^{0,t,s}$
will be regarded as free parameters and determined in this work.
Up to the second Gegenbauer terms, the Legendre polynomial $P_3(2\zeta-1)$ also
contributes. However, more unknown form factors will be introduced,
and currently available data are not sufficient for their extraction.

The time-like pion form factor $F_\pi(w^2)$ has attracted considerable
theoretical effort~\cite{plb412-382,jhep9805-014,jhep0203-046,epjc27-587,
prd63-093005+npps121-179,prd77-054015,JHEP1102-109+0408-042,Achasov,epjc71-1632,
prd83-074004,epjc72-1848+73-2453,plb715-170,prd88-093002,epjc73-2528,plb742-55}
and been measured with high precision by the CMD-2~\cite{plb527-161,plb578-285,plb648-28},
KLOE~\cite{plb606-12,plb670-285,plb700-102,plb720-336},
BaBar~\cite{prl103-231801,prd86-032013}, BESIII~\cite{plb753-629},
ALEPH~\cite{zpc76-15,pr421-191}, CLEO~\cite{prl72-3762,prd61-112002}, and
Belle~\cite{prd78-072006} Collaborations. The $\rho$ meson dominance model for
$F_\pi$ has been established in Ref.~\cite{pr124-953}. Guaranteed by the Watson 
theorem~\cite{Watson-1952}, strong interactions between the $\rho$ meson and the 
pion pair, including elastic rescattering of the two pions, can be factorized 
into $F_\pi$. In experimental investigations of three-body hadronic $B$ meson decays, 
the $\rho$ resonant contribution is usually parametrized as the Gounaris-Sakurai (GS)
model~\cite{prl21-244} based on the Breit-Wigner (BW) function~\cite{BW-model}.
Taking into account the $\rho$-$\omega$ interference and excited-state
contributions, we write $F_\pi$ as a coherent sum~\cite{prd86-032013}
\begin{eqnarray}
F_\pi(w^2)=\left[{\rm GS}_\rho(w^2,m_\rho,\Gamma_\rho)
                \frac{1+c_\omega{\rm BW}_\omega(w^2,m_\omega,\Gamma_\omega)}
              {1+c_\omega}+\sum c_i{\rm GS}_i(w^2,m_i,\Gamma_i)\right]
              \left(1+\sum c_i\right)^{-1},
\label{eqn-fpi}
\end{eqnarray}
with $i=\rho^\prime(1450)$, $\rho^{\prime\prime}(1700)$ and  
$\rho^{\prime\prime\prime}(2254)$. The explicit 
expressions of the auxiliary functions GS and BW in Eq.~(\ref{eqn-fpi}) 
are referred to Refs.~\cite{prl21-244,prd86-032013}. The inputs for the
masses $m$ and widths $\Gamma$ of $\rho^\prime$, $\rho^{\prime\prime}$, and 
$\rho^{\prime\prime\prime}$, and for the complex parameters $c$ can be found 
in Ref.~\cite{prd86-032013}. We have $c_\omega=0$ for a charged $\rho$ meson, 
because of no interference between it and a $\omega$ meson. Note that the 
Gegenbauer moments $a_2^{0,t,s}$ are the same for all the resonant states
$\rho$, $\rho^{\prime}$, $\rho^{\prime\prime}$,... in the above parametrization 
of the two-pion distribution amplitudes.

The quasi-two-body decays $B\to K\rho \to K\pi\pi$ can be also analyzed in an
alternative approach based on 
two-body decays: the quark pair $q\bar q$ from a hard decay kernel forms the $\rho$
meson, followed by its BW propagator, and then by
the $\rho\to\pi\pi$ transition with the strength $g_{\rho\pi\pi}$.
The equivalence between the framework with the $\rho$ meson propagator and the present
one with the two-pion distribution amplitudes hints the relation,
\begin{eqnarray}
F^\rho_\pi(w^2)\approx\frac{g_{\rho\pi\pi} w f_\rho}{ D_\rho(w^2)},\label{frho}
\end{eqnarray}
where $F^\rho_\pi$ represents the $\rho$ component of Eq.~(\ref{eqn-fpi}), $f_\rho$ is
the $\rho$ meson decay constant, and
$D_\rho$ is the denominator of the BW function for the $\rho$ resonance.
We have the similar relations for the $\rho$ components in the other two
form factors, $F^\rho_{s,t}(w^2)\approx g_{\rho\pi\pi} w f_\rho^T/D_\rho(w^2)$, in which
the decay constant $f_\rho^T$ normalizes the twist-3
$\rho$ meson distribution amplitudes. Due to the dominance of the $\rho$ resonant
contributions to the time-like form factors~\cite{prd86-032013},
it is legitimate to postulate the approximation
$F_{s,t}(w^2)\approx (f_\rho^T/f_\rho) F_\pi(w^2)$.

\begin{figure}[tbp]
\centerline{\epsfxsize=14cm \epsffile{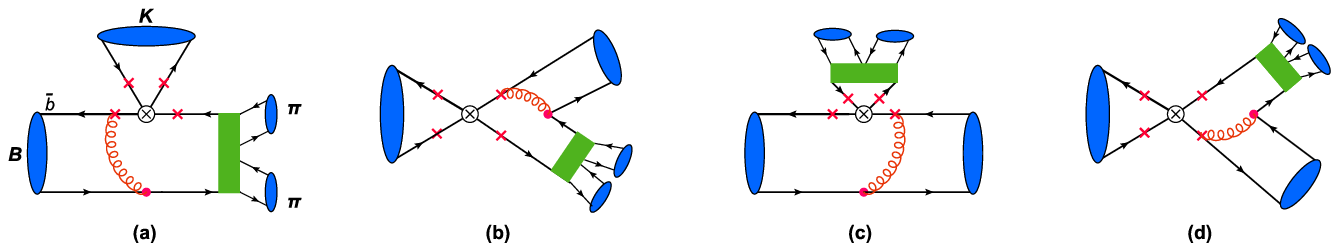}}
\caption{Typical Feynman diagrams for the quasi-two-body decays $B\to K\rho\to K\pi\pi$,
in which the symbol $\otimes$ stands for the weak vertex, $\times$ denotes
possible attachments of hard gluons, and the green rectangle represents intermediate
states.}
\label{fig-feyndiag}
\end{figure}

The amplitude ${\mathcal A}$ for the quasi-two-body decays
$B\to K\rho \to K\pi\pi$ in the PQCD approach is, according to
Fig.~\ref{fig-feyndiag}, given by~\cite{plb561-258,prd70-054006}
\begin{eqnarray}
{\mathcal A}=\phi_B\otimes H\otimes\phi_{K}\otimes \phi_{\pi\pi}^{I=1}\;,
\end{eqnarray}
where the hard kernel $H$ contains only one hard gluon exchange at leading order 
in the strong coupling $\alpha_s$ as in the two-body formalism, 
the symbol $\otimes$ means convolutions in parton momenta, and the $B$ meson (kaon,
two-pion) distribution amplitude $\phi_B$ ($\phi_{K}$, $\phi_{\pi\pi}^{I=1}$) absorbs
nonperturbative dynamics in the decay processes. We then have their
differential branching fractions~\cite{PDG-2014}
\begin{eqnarray}
\frac{d{\mathcal B}}{dw^2}=\tau_B\frac{|\overrightarrow{p_\pi}||\overrightarrow{p_K}|}
{32\pi^3m^3_B}|{\mathcal A}|^2\;,
\label{eqn-bf}
\end{eqnarray}
$\tau_B$ being the $B$ meson mean lifetime. The magnitudes of the
pion and kaon momenta, $|\overrightarrow{p_\pi}|$ and $|\overrightarrow{p_K}|$,
are written, in the center-of-mass frame of the pion pair, as
\begin{eqnarray}
 |\overrightarrow{p_\pi}|=\frac12\sqrt{w^2-4m^2_{\pi}}\;, \quad~~
 |\overrightarrow{p_K}|=\frac{1}{2}
  \sqrt{\left[\left(m^2_{B}-m_{K}^2\right)^2 -2\left(m^2_{B}+m_{K}^2\right)w^2+w^4\right]/w^2}\;,
\end{eqnarray}
with the pion mass $m_\pi$ and the kaon mass $m_K$.
The $B\to K\rho \to K\pi\pi$ decay amplitudes 
${\mathcal A}$ are collected in the Appendix,
which are similar to those in Ref.~\cite{prd74-094020} for the two-body $B$ meson
decay into a pseudoscalar meson and a vector meson.

\section{RESULTS}

For the numerical study, we adopt the inputs (in units of
GeV)~\cite{PDG-2014}
\begin{eqnarray}
\Lambda^{(f=4)}_{ \overline{MS} }&=&0.250, \quad m_{B^{\pm,0}}=5.280, \quad
  m_{K^\pm}=0.494, \quad m_{K^0}=0.498, \nonumber\\
m_{\pi^\pm}&=&0.140, \quad m_{\pi^0}=0.135, \quad~~ m_\rho=0.775, \quad~~~ \Gamma_\rho=0.149,
\end{eqnarray}
the mean lifetimes $\tau_{B^0}=1.519\times 10^{-12}$~s and $\tau_{B^\pm}=1.638\times 10^{-12}$~s,
and the Wolfenstein parameters from Ref.~\cite{PDG-2014}. The
decay constant $f_\rho$ has been extracted from the $\tau^\pm\to \rho^\pm\nu_\tau$ decay rate
for the charged $\rho^\pm$ meson and from $\rho^0\to e^+e^-$ for the neutral $\rho^0$ meson.
In this work we take their arithmetic average value
$f_\rho=(0.216\pm0.003)$ GeV~\cite{prd75-054004,1503-05534}. The decay constant $f^T_\rho$
has been computed in lattice QCD~\cite{prd80-054510,prd78-114509,prd68-054501,jhep0305-007},
for which we choose $f^T_\rho=0.184$ GeV~\cite{prd80-054510}. The ratio $f^T_\rho/f_\rho$ then
determines the ratios $F_{s,t}/F_\pi$ postulated in the previous section.
The $B$ meson and kaon distribution amplitudes
are the same as widely adopted in the PQCD approach~\cite{prd89-074031,
prd85-094003,prd86-114025,prd76-074018}.

\begin{table}[thb]
\begin{center}
\caption{PQCD results for the $CP$ averaged branching ratios and the direct $CP$
         asymmetries of the $B\to K\rho\to K\pi\pi$ decays. The corresponding
         data are quoted from Particle Data Group~\cite{PDG-2014}.}
\label{tab1}   
\begin{tabular}{l c c c} \hline\hline
\     ~~~~Mode       &    ~~      &   Results   & \; Data~\cite{PDG-2014}  \\  \hline
  $B^+\to K^+\rho^0\to K^+\pi^+\pi^-$\;     &$~~{\mathcal B}~(10^{-6})~~$
      &\; $3.42^{+0.78}_{-0.55}(\omega_B)^{+0.44}_{-0.39}(a^t_2)^{+0.39}_{-0.38}(m^K_0)
      ^{+0.39}_{-0.32}(a^0_2)^{+0.29}_{-0.28}(a^s_2)$\;    & \;$3.7\pm0.5$ \\
               &  ${\mathcal A}_{CP}$
      &\; $0.43^{+0.04}_{-0.05}(\omega_B)\pm0.06(a^t_2)\pm0.03(m^K_0)
      \pm0.03(a^0_2)\pm0.01(a^s_2)$\;    & \;$0.37\pm0.10$ \\
  $B^+\to K^0\rho^+\to K^0\pi^+\pi^0$\;     &$~~{\mathcal B}~(10^{-6})~~$
      &\; $7.43^{+1.92}_{-1.31}(\omega_B)^{+1.65}_{-1.42}(a^t_2)^{+0.88}_{-0.91}(m^K_0)
      ^{+0.60}_{-0.62}(a^0_2)^{+0.53}_{-0.47}(a^s_2)$\;    & \;$8.0\pm1.5$\; \\
              &  ${\mathcal A}_{CP}$
      &\; $0.15^{+0.02}_{-0.01}(\omega_B)^{+0.04}_{-0.05}(a^t_2)\pm0.01(m^K_0)
      ^{+0.01}_{-0.00}(a^0_2)\pm0.00(a^s_2)$\;    & \;$~ -0.12\pm0.17 ~$ \\
  $B^0\to K^+\rho^-\to K^+\pi^-\pi^0$\; &$~~{\mathcal B}~(10^{-6})~~$
      &\; $6.51^{+1.71}_{-1.12}(\omega_B)^{+0.58}_{-0.61}(a^t_2)^{+0.78}_{-0.77}(m^K_0)
      ^{+0.67}_{-0.64}(a^0_2)^{+0.39}_{-0.47}(a^s_2)$\;    & \;$7.0\pm0.9$\; \\
              &  ${\mathcal A}_{CP}$
      &\; $0.31^{+0.00}_{-0.01}(\omega_B)^{+0.09}_{-0.08}(a^t_2)^{+0.03}_{-0.02}(m^K_0)
      \pm0.01(a^0_2)\pm0.02(a^s_2)$\;    & \;$0.20\pm0.11$ \\
  $B^0\to K^0\rho^0\to K^0\pi^+\pi^-$\; &$~~{\mathcal B}~(10^{-6})~~$
      &\; $3.76^{+1.09}_{-0.74}(\omega_B)^{+0.73}_{-0.60}(a^t_2)^{+0.52}_{-0.47}(m^K_0)
      ^{+0.28}_{-0.25}(a^0_2)^{+0.26}_{-0.23}(a^s_2)$\;    & \;$4.7\pm0.6$\; \\
              &  ${\mathcal A}_{CP}$
      &\; $0.06^{+0.01}_{-0.02}(\omega_B)^{+0.00}_{-0.01}(a^t_2)\pm0.00(m^K_0)
      ^{+0.00}_{-0.01}(a^0_2)\pm0.00(a^s_2)$\;    & \;$-$ \\
\hline\hline
\end{tabular}
\end{center}
\end{table}

We first single out the $\rho(770)$ component of the time-like pion form factor
in Eq.~(\ref{eqn-fpi}). The fit to the data in Table I determines the Gegenbauer
moments $a^{0}_2=0.25$, $a^{s}_2=0.75$, and $a^{t}_2=-0.60$, which differ
from those in the distribution amplitudes for a longitudinally
polarized $\rho$ meson~\cite{prd65-014007,npb529-323}. The resultant
$CP$ averaged branching ratios (${\mathcal B}$) and direct $CP$ asymmetries
(${\mathcal A}_{CP}$) for the $B^+\to K^+\rho^0\to K^+\pi^+\pi^-$,
$B^+\to K^0\rho^+\to K^0\pi^+\pi^0$, $B^0\to K^+\rho^-\to K^+\pi^-\pi^0$ and
$B^0\to K^0\rho^0\to K^0\pi^+\pi^-$ modes are presented in Table~\ref{tab1}. The
theoretical uncertainties come from the variations of the
shape parameter of the $B$ meson distribution amplitude
$\omega_B=0.40\pm0.04$~GeV, $a^t_2=-0.60\pm0.20$, the chiral scale associate with
the kaon $m^K_0=1.6\pm0.1$~GeV,
$a^0_2=0.25\pm0.10$, and $a^s_2=0.75\pm0.25$. The uncertainties from
$\tau_{B^\pm}$, $\tau_{B^0}$, the Gegenbauer moments of the kaon distribution
amplitudes, and the Wolfenstein parameters in~\cite{PDG-2014} are small and have
been neglected. It is observed that the uncertainties of ${\mathcal A}_{CP}$
are much smaller than those of ${\mathcal B}$, and that the consistency between
our results and the data is satisfactory.

Examining the distributions of these branching ratios in the
pion-pair invariant mass $w$, we find that the main portion of the branching ratios
lies in the region around the pole mass of the $\rho$ resonance as expected:
the differential branching ratios of the
$B^\pm\to K^\pm\rho^0\to K^\pm\pi^+\pi^-$ decays in Fig.~2(a)
exhibit peaks at the $\rho$ meson mass.
The central values of ${\mathcal B}$ are $1.78\times10^{-6}$ and $2.46\times10^{-6}$
for the $B^+\to K^+\rho^0\to K^+\pi^+\pi^-$ decay in the ranges of $w$,
$[m_\rho-0.5\Gamma_\rho, m_\rho+0.5\Gamma_\rho]$ and
$[m_\rho-\Gamma_\rho, m_\rho+\Gamma_\rho]$, respectively, which amount to $52\%$ and
$72\%$ of ${\mathcal B}=3.42\times10^{-6}$ in Table~\ref{tab1}.
The branching fraction $3.27\times10^{-6}$ is accumulated in the range
$[2m_\pi, 1.5~{\rm GeV}]$ for this mode. Figure~2(b) displays the differential
distributions of ${\mathcal A}_{CP}$ for the four
$B\to K\rho\to K\pi\pi$ modes, in which a falloff of
${\mathcal A}_{CP}$ with $w$ is seen for
$B^+\to K^+\rho^0\to K^+\pi^+\pi^-$, $B^+\to K^0\rho^+\to K^0\pi^+\pi^0$,
and $B^0\to K^+\rho^-\to K^+\pi^-\pi^0$.
It implies that the direct $CP$ asymmetries in the above three quasi-two-body
decays, if calculated as the two-body decays $B\to K\rho$
with the $\rho$ resonance mass being fixed to $m_\rho$, may be overestimated.
The ascent of the differential distribution of ${\mathcal A}_{CP}$ with $w$
for $B^0\to K^0\rho^0\to K^0\pi^+\pi^-$ implies that its direct $CP$ asymmetry,
if calculated in the two-body formalism, may be underestimated .

\begin{figure}[tbp]
\centerline{\epsfxsize=7.5cm \epsffile{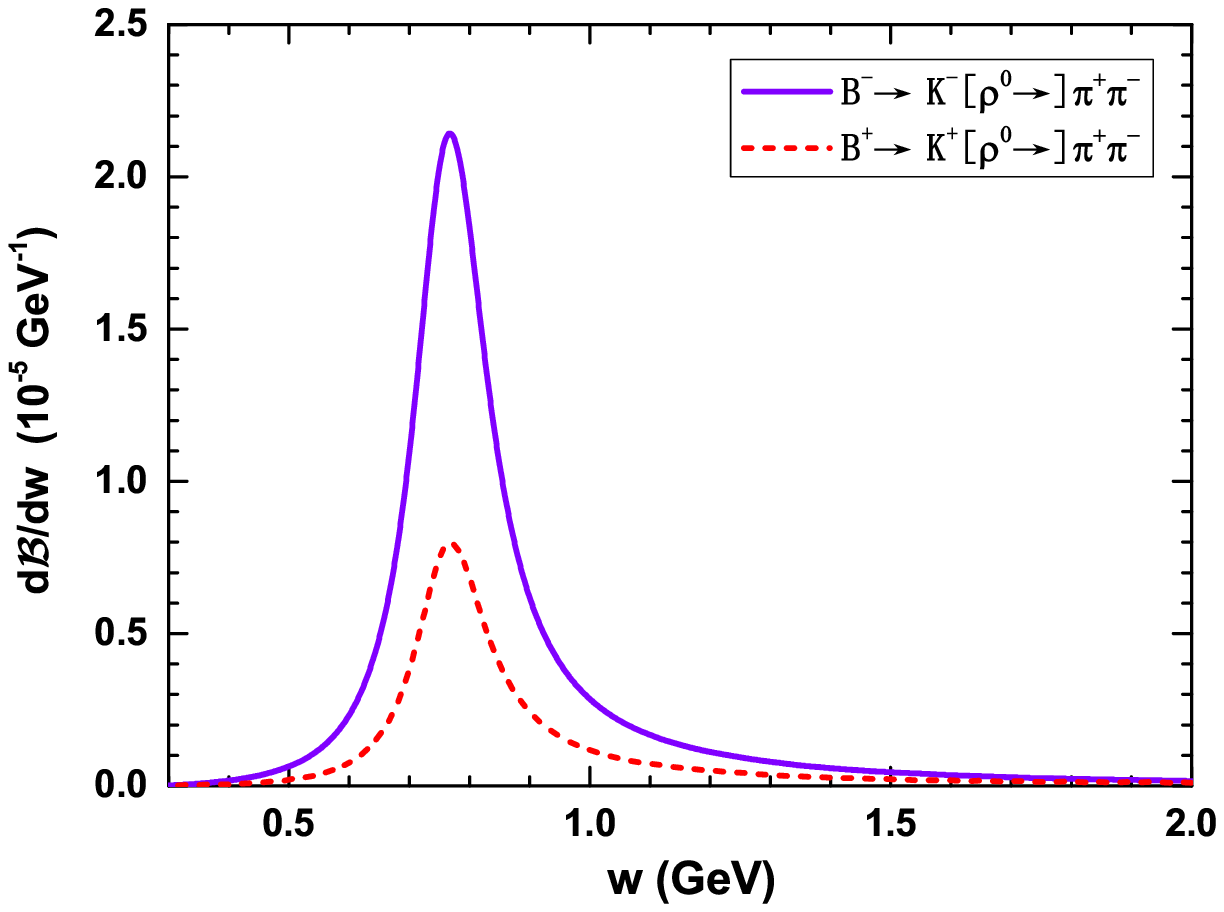}
            \epsfxsize=7.5cm \epsffile{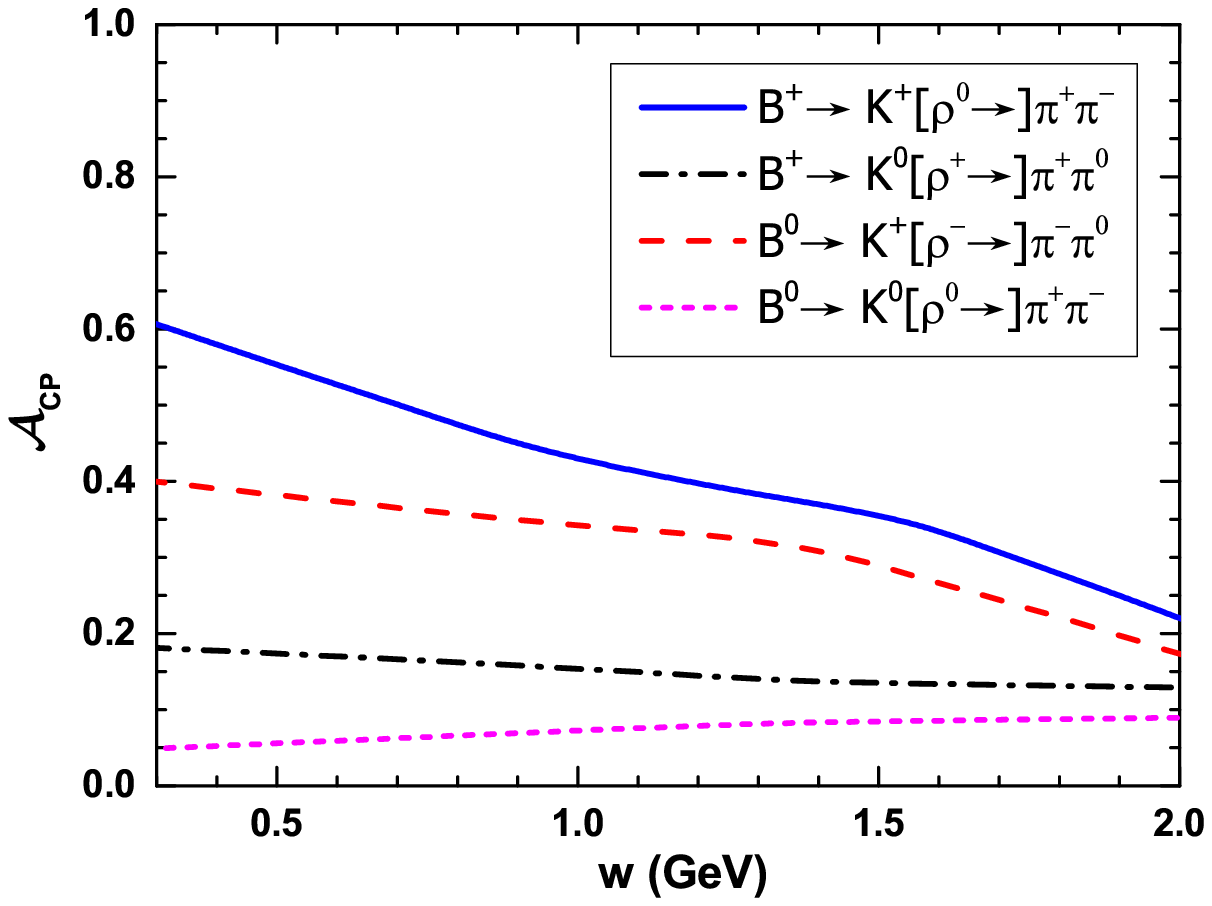}}
\vspace{-0.3cm}
  {\scriptsize\bf (a)\hspace{7.2cm}(b)}
\caption{(a) differential branching ratios for the $B^\pm\to K^\pm\rho^0\to K^\pm\pi^+\pi^-$ decays,
         and (b) differential distributions of ${\mathcal A}_{CP}$ in $w$ 
         for the $B\to K\rho \to K\pi\pi$ decays.}
\label{fig-dis-br}
\end{figure}

To verify the above observation, we treat the $B\to K\rho\to K\pi\pi$
modes as the two-body decays $B\to K\rho$ in the PQCD approach~\cite{prd74-094020} 
by imposing the replacement $\eta\to r^2_\rho$ for the momenta in
Eqs.~(\ref{def-pp3}) and (\ref{def-pp4}), with the mass ratio $r_\rho=m_\rho/m_B$.
Employing the same Gegenbauer moments $a^{0,t,s}_2$ for the 
$\rho$ meson distribution amplitudes, we obtain
\begin{eqnarray}
B^+\to K^+\rho^0 &&  \left\{ \begin{array}{ll}
 {\mathcal B}=(3.52^{+0.67}_{-0.45}(\omega_B)^{+0.40}_{-0.34}(a^t_2)^{+0.42}_{-0.38}(m^K_0)
      ^{+0.47}_{-0.43}(a^0_2)^{+0.25}_{-0.24}(a^s_2))\times 10^{-6}\;,   \\
 {\mathcal A}_{CP}= 0.55^{+0.02}_{-0.04}(\omega_B)^{+0.09}_{-0.08}(a^t_2)\pm0.03(m^K_0)
      ^{+0.00}_{-0.01}(a^0_2)\pm0.01(a^s_2)\;,   \\
                             \end{array} \right. \label{result=k+r0}\\
B^+\to K^0\rho^+ &&  \left\{ \begin{array}{ll}
 {\mathcal B}=(7.66^{+1.79}_{-1.19}(\omega_B)^{+1.69}_{-1.44}(a^t_2)^{+1.04}_{-0.95}(m^K_0)
      ^{+0.84}_{-0.73}(a^0_2)^{+0.43}_{-0.41}(a^s_2))\times 10^{-6}\;,   \\
 {\mathcal A}_{CP}= 0.22\pm0.03(\omega_B)^{+0.03}_{-0.05}(a^t_2)\pm0.01(m^K_0)
      \pm0.00(a^0_2)\pm0.00(a^s_2)\;,   \\
                             \end{array} \right. \label{result=k0r+}\\
B^0\to K^+\rho^- &&  \left\{ \begin{array}{ll}
 {\mathcal B}=(6.92^{+1.58}_{-1.04}(\omega_B)^{+0.67}_{-0.53}(a^t_2)^{+0.86}_{-0.81}(m^K_0)
      ^{+0.91}_{-0.80}(a^0_2)^{+0.42}_{-0.40}(a^s_2))\times 10^{-6}\;,   \\
 {\mathcal A}_{CP}= 0.34^{+0.00}_{-0.01}(\omega_B)^{+0.13}_{-0.12}(a^t_2)^{+0.03}_{-0.02}(m^K_0)
      ^{+0.01}_{-0.02}(a^0_2)^{+0.01}_{-0.02}(a^s_2)\;.   \\
                             \end{array} \right. \label{result=k+r-}\\
B^0\to K^0\rho^0 &&  \left\{ \begin{array}{ll}
 {\mathcal B}=(4.01^{+1.07}_{-0.71}(\omega_B)^{+0.70}_{-0.63}(a^t_2)^{+0.55}_{-0.50}(m^K_0)
      ^{+0.40}_{-0.35}(a^0_2)\pm0.19(a^s_2))\times 10^{-6}\;,   \\
 {\mathcal A}_{CP}= 0.04\pm0.01(\omega_B)\pm0.00(a^t_2)\pm0.00(m^K_0)
      ^{+0.00}_{-0.01}(a^0_2)\pm0.00(a^s_2)\;.   \\
                             \end{array} \right. \label{result=k0r0}
\end{eqnarray}
The comparison of Table~\ref{tab1} with Eqs.~(\ref{result=k+r0})-(\ref{result=k0r0})
confirms that the branching ratios of the four quasi-two-body modes in the three-body
and two-body frameworks are close to each other. The tiny distinction between
them suggests that the PQCD approach is a consistent theory for exclusive hadronic $B$ meson
decays. The total ${\mathcal A}_{CP}$ for the decays $B^+\to K^+\rho^0\to K^+\pi^+\pi^-$,
$B^+\to K^0\rho^+\to K^0\pi^+\pi^0$, and $B^0\to K^+\rho^-\to K^+\pi^-\pi^0$
in Table~\ref{tab1}, compared with the corresponding values
in Eqs.~(\ref{result=k+r0})-(\ref{result=k+r-}), have been, as explained above,
moderated by the finite width of the $\rho$ resonance appearing in the time-like
form factor $F_\pi$. Because ${\mathcal A}_{CP}$ in Table~\ref{tab1} agree better
with the data, it may be more appropriate to treat $B\to K\rho$ as
three-body decays.

\begin{table}[thb]
\begin{center}
\caption{PQCD predictions for the $CP$ averaged branching ratios and the direct $CP$
asymmetries of the $B\to K\rho^\prime\to K\pi\pi$ decays.}
\label{tab2}
\begin{tabular}{l c c} \hline\hline     
\     ~~~~Mode       &    ~~      &   Results      \\  \hline
  $B^+\to K^+\rho^{\prime0}\to K^+\pi^+\pi^-$\;     &$~~{\mathcal B}~(10^{-7})~~$
      &\; $4.32^{+1.17}_{-0.99}(\omega_B)^{+0.81}_{-0.79}(a^t_2)^{+0.59}_{-0.64}(a^s_2)
            ^{+0.40}_{-0.46}(m^K_0)^{+0.13}_{-0.17}(a^0_2)$\;     \\
               &  ${\mathcal A}_{CP}$
      &\; $0.32^{+0.06}_{-0.04}(\omega_B)\pm0.03(a^t_2)^{+0.01}_{-0.02}(a^s_2)
            ^{+0.02}_{-0.01}(m^K_0)\pm0.01(a^0_2)$\;     \\
  $B^+\to K^0\rho^{\prime+}\to K^0\pi^+\pi^0$\;     &$~~{\mathcal B}~(10^{-7})~~$
      &\; $10.37^{+3.72}_{-2.36}(\omega_B)^{+3.14}_{-2.71}(a^t_2)^{+1.26}_{-1.03}(a^s_2)
            ^{+1.13}_{-0.92}(m^K_0)^{+0.42}_{-0.37}(a^0_2)$\;     \\
              &  ${\mathcal A}_{CP}$
      &\; $0.12\pm0.02(\omega_B)^{+0.02}_{-0.01}(a^t_2)^{+0.03}_{-0.02}(a^s_2)
            \pm0.01(m^K_0)\pm0.01(a^0_2)$\;     \\
  $B^0\to K^+\rho^{\prime-}\to K^+\pi^-\pi^0$\; &$~~{\mathcal B}~(10^{-7})~~$
      &\; $7.61^{+2.37}_{-1.90}(\omega_B)^{+1.32}_{-1.03}(a^t_2)^{+1.17}_{-0.88}(a^s_2)
            ^{+0.86}_{-0.75}(m^K_0)^{+0.26}_{-0.22}(a^0_2)$\;     \\
              &  ${\mathcal A}_{CP}$
      &\; $0.27^{+0.02}_{-0.01}(\omega_B)\pm0.06(a^t_2)^{+0.00}_{-0.01}(a^s_2)
            \pm0.02(m^K_0)\pm0.01(a^0_2)$\;     \\
  $B^0\to K^0\rho^{\prime0}\to K^0\pi^+\pi^-$\; &$~~{\mathcal B}~(10^{-7})~~$
      &\; $4.84^{+1.82}_{-1.32}(\omega_B)^{+1.11}_{-1.05}(a^t_2)\pm0.50(a^s_2)
            ^{+0.48}_{-0.46}(m^K_0)^{+0.14}_{-0.16}(a^0_2)$\;     \\
              &  ${\mathcal A}_{CP}$
      &\; $0.08^{+0.00}_{-0.01}(\omega_B)^{+0.02}_{-0.00}(a^t_2)\pm0.01(a^s_2)
            \pm0.01(m^K_0)\pm0.01(a^0_2)$\;     \\
\hline\hline
\end{tabular}
\end{center}
\end{table}

\begin{figure}[tbp]
\centerline{\epsfxsize=7.5cm \epsffile{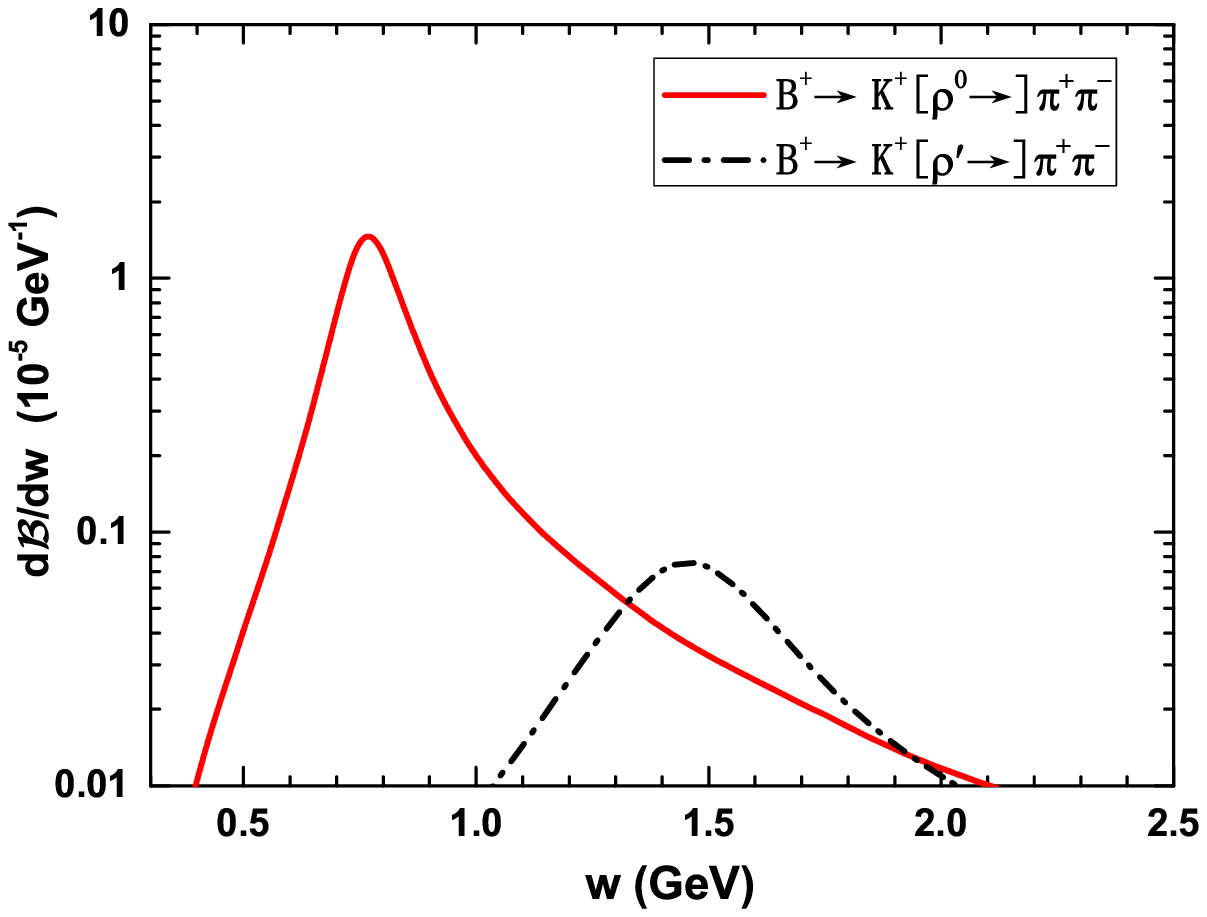}}
\vspace{-0.5cm}
\caption{$CP$ averaged differential branching ratios for the decays
$B^+\to K^+\rho^0\to K^+\pi^+\pi^-$ and $B^+\to K^+\rho^{\prime0}\to K^+\pi^+\pi^-$.}
\label{fig-dep}
\end{figure}

The parametrization of the time-like pion form factor in Eq.~(\ref{eqn-fpi}) also
allows to single out the the $\rho^\prime(1450)$ component. Adopting the two-pion
distribution amplitudes in Eqs.~(\ref{def-DA0})-(\ref{def-DAs}), we derive
the $CP$ averaged branching ratios and the direct $CP$ asymmetries for the decays
$B^+\to K^+\rho^{\prime 0}\to K^+\pi^+\pi^-$, $B^+\to K^0\rho^{\prime+}\to K^0\pi^+\pi^0$,
$B^0\to K^+\rho^{\prime-}\to K^+\pi^-\pi^0$, and $B^0\to K^0\rho^{\prime0}\to K^0\pi^+\pi^-$
listed in Table II, whose errors have the same sources as in Table I.
We compare the differential branching ratios for the
$B^+\to K^+\rho^0\to K^+\pi^+\pi^-$ and $B^+\to K^+\rho^{\prime0}\to K^+\pi^+\pi^-$
decays in Fig.~\ref{fig-dep}, whose difference is mainly governed by the
corresponding BW functions. All these predictions
can be confronted with data in the future.

To extract the branching ratios for the two-body decays $B\to K\rho^\prime$
from the quasi-two-body ones $B\to K\rho^\prime\to K\pi\pi$, we need
the branching fraction for $\rho^\prime\to \pi\pi$, which is inferred from
the ratio of the widths, $\Gamma_{\pi\pi}/\Gamma_{\rho^\prime}$. The width
$\Gamma_{\pi\pi}$ for $\rho^\prime\to \pi\pi$ was evaluated in the
Nambu-Jona-Lasinio quark model, and found to be
$22$~MeV~\cite{ijmpa13-5443}, consistent with $17\sim25$~MeV
obtained from the $e^+e^-$ annihilation data~\cite{zpc62-455}.
Taking $\Gamma_{\rho^\prime}=0.311\pm0.062$ GeV~\cite{zpc62-455}, we get
the branching fraction ${\mathcal B}(\rho^\prime\to\pi\pi)=4.56\%\sim10.0\%$.
The $\rho^\prime\to \pi\pi$ branching fraction can be also estimated from
the relation~\cite{EPJC2-269}
\begin{eqnarray}
\Gamma_{\rho^\prime\to\pi\pi}=\frac{g^2_{\rho^\prime\pi\pi}}{6\pi}
\frac{|\overrightarrow{p_\pi}(m^2_{\rho^\prime})|^3}{m^2_{\rho^\prime}}.
\label{bf}
\end{eqnarray}
The coupling $g_{\rho^\prime\pi\pi}$ is read off the $\rho^\prime$
component of the time-like form factor $F_\pi$ in Eq.~(\ref{eqn-fpi}) according to
$F^{\rho^\prime}_\pi(w^2)\approx g_{\rho^\prime\pi\pi} w f_{\rho^\prime}/D_{\rho^\prime}(w^2)$
at $w=m_{\rho^\prime}$, which is similar to Eq.~(\ref{frho}) for the $\rho$ component.
We adopt the decay constant $f_{\rho^\prime}=0.185^{+0.030}_{-0.035}$~GeV
resulting from the data $\Gamma_{\rho^\prime\to e^+e^-}=1.6\sim3.4$
keV~\cite{zpc62-455}, which agrees with $f_{\rho^\prime}=(0.182\pm0.005)$~GeV
from the perturbative analysis in the large-$N_c$ limit~\cite{prd77-116009},
$f_{\rho^\prime}=(0.186\pm0.014)$~GeV from the double-pole QCD sum
rules~\cite{1205-6793}, and $f_{\rho^\prime}=0.128$~GeV from the relativistic
constituent quark model~\cite{prd60-094020}. Equation~(\ref{bf}) then yields
${\mathcal B}(\rho^\prime\to\pi\pi)=10.04^{+5.23}_{-2.61}\%$,
compatible with ${\mathcal B}(\rho^\prime\to\pi\pi)=4.56\%\sim10.0\%$
from the width ratio $\Gamma_{\pi\pi}/\Gamma_{\rho^\prime}$.

With ${\mathcal B}(\rho^\prime\to\pi\pi)=10.04\%$, we extract the $B\to K\rho^{\prime}$ 
branching ratios from Table II (in units of $10^{-6}$),
\begin{eqnarray}
\label{br-2body-0p}
{\mathcal B}(B^+\to K^+\rho^{\prime0})&=& 4.30^{+1.16}_{-0.99}(\omega_B)^{+0.80}_{-0.79}(a^t_2)
            ^{+0.59}_{-0.64}(a^s_2)^{+0.40}_{-0.45}(m^K_0)^{+0.13}_{-0.17}(a^0_2) \;,\\
{\mathcal B}(B^+\to K^0\rho^{\prime+})&=& 10.33^{+3.71}_{-2.35}(\omega_B)^{+3.13}_{-2.70}(a^t_2)
            ^{+1.26}_{-1.03}(a^s_2)^{+1.13}_{-0.92}(m^K_0)^{+0.41}_{-0.37}(a^0_2)\;,\\
\label{br-2body-Rp}
{\mathcal B}(B^0\to K^+\rho^{\prime-})&=& 7.57^{+2.36}_{-1.89}(\omega_B)^{+1.31}_{-1.03}(a^t_2)
            ^{+1.16}_{-0.87}(a^s_2)^{+0.86}_{-0.74}(m^K_0)^{+0.26}_{-0.22}(a^0_2)\;,\\
{\mathcal B}(B^0\to K^0\rho^{\prime0})&=& 4.82^{+1.82}_{-1.31}(\omega_B)^{+1.11}_{-1.04}(a^t_2)
            \pm0.50(a^s_2)^{+0.47}_{-0.46}(m^K_0)^{+0.14}_{-0.16}(a^0_2)\;.
\end{eqnarray}
Note that the data ${\mathcal B}(B^0\to K^+\rho^{\prime-})=(2.4\pm1.0\pm0.6)\times10^{-6}$ from
BaBar~\cite{prd83-112010} by assuming
${\mathcal B}(\rho^\prime\to\pi\pi)\approx 100\%$ is much larger
than Eq.~(\ref{br-2body-Rp}) based on ${\mathcal B}(\rho^\prime\to\pi\pi)=4.56\%\sim10.0\%$
or $10.04^{+5.23}_{-2.61}\%$, and the data
${\mathcal B}(B^0\to K^+\rho^-)=(6.6\pm0.5\pm0.8)\times10^{-6}$ in Ref~\cite{prd83-112010}.

The branching ratios and the direct CP asymmetries of the quasi-two-body decays
$B\to K(\omega,\rho^{\prime\prime}, \rho^{\prime\prime\prime})\to K\pi\pi$
can be predicted by singling out the corresponding components in the time-like form
factor $F_\pi$ in principle, since the Gegenbauer moments of the $P$-wave two-pion
distribution amplitudes have been determined. This is a merit of our PQCD
formalism for three-body hadronic $B$ meson decay. Besides, we can extract, for example,
the $B\to K\omega$ branching ratios from the predictions for the
$B\to K\omega\to K\pi\pi$ modes, given the $\omega\to\pi\pi$ branching
fraction. We will leave the above observables to future studies.

\section{CONCLUSION}

In this paper we have applied the PQCD approach to the quasi-two-body decays
$B\to K\rho\to K\pi\pi$, which were analyzed in both three-body and two-body
factorization formalisms. In the former strong dynamics between the $P$-wave resonances 
and the pion pair, including two-pion final-state interactions, is parametrized 
into the two-pion distribution amplitudes. The advantage of this approach is that the
time-like pion form factor $F_\pi$ involved in the two-pion distribution amplitudes 
accommodates both resonant and nonresonant contributions.
Inputting $F_\pi$ extracted from the $e^+e^-$ annihilation
data, we have calculated the branching ratios and the direct $CP$ asymmetries
of the $B\to K\rho\to K\pi\pi$ modes, whose
agreement with the data was achieved by tuning the Gegenbauer moments of the
$P$-wave two-pion distribution amplitudes. The consistency between
the three-body and two-body analyses of the $B\to K\rho \to K\pi\pi$ branching ratios 
was verified, which supports the PQCD approach
to exclusive hadronic $B$ meson decays. The comparison to the results
from the two-body framework indicates that the direct $CP$ asymmetries
of the $B\to K\rho\to K\pi\pi$ modes have been moderated by the finite width of the
$\rho$ resonance, and become closer to the data. It suggests that the
three-body framework is more appropriate for studying quasi-two-body
hadronic $B$ meson decays.

The contribution from the $\rho^\prime$ intermediate state was
simply singled out from the given time-like form factor $F_\pi$ in our formalism. 
Using the determined Gegenbauer moments of the $P$-wave two-pion distribution amplitudes, 
we have predicted the branching ratios and the direct $CP$ asymmetries of the 
$B\to K\rho^\prime\to K\pi\pi$ channels, and compared their differential branching ratios
with the $B\to K\rho\to K\pi\pi$ ones. With the estimated $\rho^\prime\to\pi\pi$ branching 
fraction, the two-body $B\to K\rho^\prime$ branching ratios have been extracted from 
the results for the $B\to K\rho^\prime\to K\pi\pi$ decays. All these predictions can be
confronted with future data. The same framework is
applicable straightforwardly to other channels
$B\to K(\omega,\rho^{\prime\prime}, \rho^{\prime\prime\prime})\to K\pi\pi$
in principle. Moreover, with both the $S$-wave and $P$-wave distribution amplitudes
being ready, we will proceed to predictions for differential branching ratios and
direct $CP$ asymmetries of three-body hadronic $B$ meson decays in various localized
regions of two-pion phase space in a forthcoming paper.

\begin{acknowledgments}
We thank Prof. H.-Y. Cheng for valuable discussions. This work was supported in part by National
Science Foundation of China under Grant No. 11547038, and by the Ministry of Science and Technology
of R.O.C. under Grant No. MOST-104-2112-M-001-037-MY3.
\end{acknowledgments}

\appendix

\section{DECAY AMPLITUDES}

The quasi-two-body $B\to K\rho\to K\pi\pi$ decay amplitudes are given, in the PQCD
approach, by
\begin{eqnarray}
{\mathcal A}\big(B^+\to K^+[\rho^0\to]\pi^+\pi^-\big)&=&\frac{G_F}{2}\bigg\{ V^*_{ub}V_{us}\bigg[
  \left(\frac{C_1}{3}+C_2\right)\left(F^{LL}_{T\rho}+F^{LL}_{A\rho}\right)
  +C_1 \left( M^{LL}_{T\rho} + M^{LL}_{A\rho} \right) +\left(C_1+\frac{C_2}{3}\right)F^{LL}_{TK}
\nonumber\\
 &+&C_2 M^{LL}_{TK}\bigg]
     -V^*_{tb}V_{ts}\bigg[ \left(\frac{C_3}{3}+C_4+\frac{C_9}{3}+C_{10}\right)
        \left(F^{LL}_{T\rho}+F^{LL}_{A\rho}\right)  \nonumber\\
 &+&\left(\frac{C_5}{3}+C_6+\frac{C_7}{3}+C_{8}\right)\left(F^{SP}_{T\rho}+F^{SP}_{A\rho}\right)
    +\left( C_3 + C_9 \right)\left(M^{LL}_{T\rho} +M^{LL}_{A\rho}\right) \nonumber\\
 &+&\left(C_5 + C_7 \right) \left(M^{LR}_{T\rho}+ M^{LR}_{A\rho} \right)
    +\left(\frac{3C_9}{2}+\frac{C_{10}}{2}\right)F^{LL}_{TK}
    +\left(\frac{3C_7}{2}+\frac{C_{8}}{2}\right)F^{LL}_{TK} \nonumber\\
 &+&\frac{3C_{10}}{2} M^{LL}_{TK}  + \frac{3C_{8}}{2} M^{SP}_{TK} \bigg] \bigg\}
 \;,\\
\label{eqn=K+r0}
{\mathcal A}\big(B^+\to K^0[\rho^+\to]\pi^+\pi^0\big)&=&\frac{G_F}{\sqrt2}\bigg\{
 V^*_{ub}V_{us}\bigg[\left(\frac{C_1}{3}+C_2\right)F^{LL}_{A\rho}+ C_1 M^{LL}_{A\rho}  \bigg]
 -V^*_{tb}V_{ts}\bigg[\left(\frac{C_3}{3}+C_4-\frac{C_9}{6}-\frac{C_{10}}{2}\right)F^{LL}_{T\rho}
\nonumber\\
 &+&\left(\frac{C_5}{3}+C_6-\frac{C_7}{6}-\frac{C_8}{2}\right)F^{SP}_{T\rho}
    +\left(C_3 -\frac{C_9}{2} \right)M^{LL}_{T\rho} +\left(C_5 -\frac{C_7}{2} \right) M^{LR}_{T\rho}
\nonumber\\
 &+&\left(\frac{C_3}{3}+C_4+\frac{C_9}{3}+C_{10}\right)F^{LL}_{A\rho}
         +\left(\frac{C_5}{3}+C_6+\frac{C_7}{3}+C_{8}\right)F^{SP}_{A\rho}
         +\left( C_3 + C_9 \right)M^{LL}_{A\rho} \nonumber\\
 &+&\left(C_5 + C_7 \right) M^{LR}_{A\rho}   \bigg] \bigg\}
\;,\\
\label{eqn=K0r+}
{\mathcal A}\big(B^0\to K^+[\rho^-\to]\pi^0\pi^-\big)&=&\frac{G_F}{\sqrt2}\bigg\{ V^*_{ub}V_{us}\bigg[
\left(\frac{C_1}{3}+C_2\right)F^{LL}_{T\rho} +C_1  M^{LL}_{T\rho} \bigg]
  -V^*_{tb}V_{ts}\bigg[ \left(\frac{C_3}{3}+C_4+\frac{C_9}{3}+C_{10}\right)F^{LL}_{T\rho} \nonumber\\
 &+&\left(\frac{C_5}{3}+C_6+\frac{C_7}{3}+C_{8}\right)F^{SP}_{T\rho}
   +\left( C_3 + C_9 \right)M^{LL}_{T\rho} +\left(C_5 + C_7 \right) M^{LR}_{T\rho} \nonumber\\
 &+&\left(\frac{C_3}{3}+C_4-\frac{C_9}{6}-\frac{C_{10}}{2}\right)F^{LL}_{A\rho}
    +\left(\frac{C_5}{3}+C_6-\frac{C_7}{6}-\frac{C_{8}}{2}\right)F^{SP}_{A\rho}
    +\left(C_3 -\frac{C_{9}}{2} \right) M^{LL}_{A\rho} \nonumber\\
 &+&\left(C_5 -\frac{C_{7}}{2} \right) M^{LR}_{A\rho}  \bigg\}
\;,\\
\label{eqn=K+r-}
{\mathcal A}\big(B^0\to K^0[\rho^0\to]\pi^+\pi^-\big)&=& -\frac{1}{\sqrt2}
          \big\{{\mathcal A}\big(B^+\to K^0[\rho^+\to]\pi^+\pi^0\big)
          +{\mathcal A}\big(B^0\to K^+[\rho^-\to]\pi^0\pi^-\big) \big\}\nonumber\\
          &+&{\mathcal A}\big(B^+\to K^+[\rho^0\to]\pi^+\pi^-\big)\;,
\end{eqnarray}
in which $G_F$ is the Fermi coupling constant, $V$'s are the Cabibbo-Kobayashi-Maskawa matrix
elements, and the amplitudes $F$ ($M$) denote the factorizable (nonfactorizable) 
contributions. It should be understood that the Wilson coefficients $C$ and the amplitudes
$F$ and $M$ appear in convolutions in momentum fractions and impact parameters $b$.
With the ratio $r=m^K_{0}/m_B$, the amplitudes from Fig.~1(a) are written as
\begin{eqnarray}
F^{LL}_{T\rho} &=& 8\pi C_F m^4_B f_K\int dx_B dz\int b_B db_B b db \phi_B(x_B,b_B)(1-\eta)
                 \big\{\big[\sqrt{\eta}(1-2z)(\phi^s+\phi^t)+(1+z)\phi^0 \big]\nonumber\\
 &\times&E_{1ab}(t_{1a})h_{1a}(x_B,z,b_B,b)
   +\sqrt{\eta}\left(2\phi^s-\sqrt\eta\phi^0 \right)E_{1ab}(t_{1b})h_{1b}(x_B,z,b_B,b) \big\}\;, \\
F^{SP}_{T\rho} &=&-16\pi C_F m^4_B r f_K \int dx_B dz\int b_B db_B b db \phi_B(x_B,b_B)
  \big\{\left[\sqrt{\eta}(2+z)\phi^s-\sqrt{\eta}z\phi^t+(1+\eta(1-2z))\phi^0 \right]\nonumber\\
 &\times& E_{1ab}(t_{1a})h_{1a}(x_B,z,b_B,b)
 +\left[2\sqrt{\eta}(1-x_B+\eta)\phi^s+(x_B-2\eta)\phi^0 \right]E_{1ab}(t_{1b})h_{1b}(x_B,z,b_B,b)\big\}\;,\\
M^{LL}_{T\rho} &=& 32\pi C_F m^4_B/\sqrt{2N_c} \int dx_B dz dx_3\int b_B db_B b_3 db_3
     \phi_B(x_B,b_B)\phi^A_K(1-\eta)\nonumber\\
&\times&\big\{\left[\sqrt{\eta}z(\phi^t-\phi^s)+((1-\eta)(1-x_3)-x_B+z\eta)\phi^0 \right]
     E_{1cd}(t_{1c})h_{1c}(x_B,z,x_3,b_B,b_3)\nonumber\\
&+&\left[z(\sqrt{\eta}(\phi^s+\phi^t)-\phi^0 )-(x_3(1-\eta)-x_B)\phi^0 \right]E_{1cd}(t_{1d})
     h_{1d}(x_B,z,x_3,b_B,b_3) \big\}\;,
\end{eqnarray}
\begin{eqnarray}
M^{LR}_{T\rho} &=& -32\pi C_F r m^4_B/\sqrt{2N_c}\int dx_B dz dx_3\int b_B db_B b_3 db_3
    \phi_B(x_B,b_B)\big\{\big[\sqrt{\eta}z(\phi^P_K-\phi^T_K)(\phi^s+\phi^t)\nonumber\\
&+&\sqrt{\eta}((1-x_3)(1-\eta)-x_B)(\phi^P_K+\phi^T_K)(\phi^s-\phi^t)
    +((1-x_3)(1-\eta)-x_B)(\phi^P_K+\phi^T_K)\phi^0+\eta z
     (\phi^P_K-\phi^T_K)\phi^0 \big]\nonumber\\
&\times& E_{1cd}(t_{1c})h_{1c}(x_B,z,x_3,b_B,b_3)
   +\big[-\sqrt{\eta}z(\phi^P_K+\phi^T_K)(\sqrt{\eta}\phi^0+(\phi^t+\phi^s))
   +(x_B-x_3(1-\eta))(\phi^P_K-\phi^T_K)\nonumber\\
&\times& (\sqrt{\eta}(\phi^s-\phi^t)+\phi^0)\big] E_{1cd}(t_{1d})h_{1d}(x_B,z,x_3,b_B,b_3)\big\}\;,
\end{eqnarray}
with the color factor $C_F=4/3$ and the kaon decay constant $f_K$.
The amplitudes from Fig.~1(b) are written as
\begin{eqnarray}
F^{LL}_{A\rho} &=& 8\pi C_F m^4_B f_B\int dz dx_3\int b db b_3 db_3 \big\{\left[2r\sqrt{\eta}\phi^P_K
    ((2-z)\phi^s+z\phi^t)-(1-\eta)(1-z)\phi^A_K\phi^0\right] E_{4ab}(t_{4a}) \nonumber\\
&\times& h_{4a}(z,x_3,b,b_3) +\big[2r\sqrt{\eta}[(1-x_3)(1-\eta)\phi^T_K
    -(1+x_3+(1-x_3)\eta)\phi^P_K]\phi^s +(x_3(1-\eta)+\eta)(1-\eta)\phi^A_K\phi^0 \big]\nonumber\\
&\times& E_{4ab}(t_{4b})h_{4b}(z,x_3,b,b_3) \big\}\;,\\
F^{SP}_{A\rho} &=&  16\pi C_F m^4_B f_B \int dz dx_3\int b db b_3 db_3
    \big\{\left[\sqrt{\eta}(1-\eta)(1-z)\phi^A_K(\phi^s+\phi^t)
    -2r(1+(1-z)\eta)\phi^P_K\phi^0 \right] E_{4ab}(t_{4a}) \nonumber\\
&\times& h_{4a}(z,x_3,b,b_3) +\left[2\sqrt{\eta}(1-\eta)\phi^A_K\phi^s - r\left(2\eta \phi^P_K
    +x_3(1-\eta)(\phi^P_K-\phi^T_K)\right)\phi^0\right]E_{4ab}(t_{4b})h_{4b}(z,x_3,b,b_3)\big\}\;,
    \nonumber\\
    \\
M^{LL}_{A\rho} &=& 32\pi C_F m^4_B/\sqrt{2N_c} \int dx_B dz dx_3\int b_B db_B b_3 db_3\phi_B(x_B,b_B)
   \big\{\big[(\eta-1)[x_3(1-\eta)+x_B +\eta(1-z)]\phi^A_K\phi^0\nonumber\\
&+& r\sqrt\eta(x_3(1-\eta)+x_B+\eta)(\phi^P_K+\phi^T_K)(\phi^s-\phi^t)+ r\sqrt\eta(1-z)
  (\phi^P_K-\phi^T_K)(\phi^s+\phi^t)+2r\sqrt\eta(\phi^P_K\phi^s+\phi^T_K\phi^t)\big]\nonumber\\
&\times& E_{4cd}(t_{4c})h_{4c}(x_B,z,x_3,b_B,b_3)+\big[(1-\eta^2)(1-z)\phi^A_K\phi^0
  +r\sqrt\eta(x_B-x_3(1-\eta)-\eta)(\phi^P_K-\phi^T_K)(\phi^s+\phi^t)\nonumber\\
&-&r\sqrt\eta(1-z)(\phi^P_K+\phi^T_K)(\phi^s-\phi^t)\big]E_{4cd}(t_{4d})
  h_{4d}(x_B,z,x_3,b_B,b_3)\big\}\;,\\
M^{LR}_{A\rho} &=& -32\pi C_F m^4_B/\sqrt{2N_c} \int dx_B dz dx_3\int b_B db_B b_3 db_3
  \phi_B(x_B,b_B)\big\{\big[\sqrt\eta(1-\eta)(1+z)\phi^A_K(\phi^s-\phi^t)+r(2-x_B\nonumber\\
&-&x_3(1-\eta))(\phi^P_K+\phi^T_K)\phi^0 +r\eta(z\phi^P_K-(2+z)\phi^T_K)\phi^0\big]
  E_{4cd}(t_{4c})h_{4c}(x_B,z,x_3,b_B,b_3) + \big[\sqrt\eta(1-\eta)(1-z)\phi^A_K \nonumber\\
&\times&(\phi^s-\phi^t)+r((x_3(1-\eta)-x_B)(\phi^P_K+\phi^T_K)
  +\eta((2-z)\phi^P_K+z\phi^T_K))\phi^0 \big]E_{4cd}(t_{4d})h_{4d}(x_B,z,x_3,b_B,b_3)\big\}\;,
  \nonumber\\
\end{eqnarray}
with the $B$ meson decay constant $f_B$.
The amplitudes from Fig.~1(c) are written as
\begin{eqnarray}
F^{LL}_{TK} &=& 8\pi C_F m^4_B \int dx_B dx_3\int b_B db_B b_3 db_3 \phi_B(x_B,b_B)
   \big\{\big[(1+x_3(1-\eta))(1-\eta)\phi^A_K + r(1-2x_3)(1-\eta)\phi^P_K  \nonumber\\
&+&r(1+\eta-2x_3(1-\eta))\phi^T_K\big]E_{2ab}(t_{2a})h_{2a}(x_B,x_3,b_B,b_3)
   +\left[x_B(\eta - 1)\eta\phi^A_K+2r(1-\eta(1-x_B))\phi^P_K\right]E_{2ab}(t_{2b}) \nonumber\\
&\times& h_{2b}(x_B,x_3,b_B,b_3)\big\}\;,\\
M^{LL}_{TK} &=& 32\pi C_F m^4_B/\sqrt{2N_c} \int dx_B dz dx_3\int b_B db_B b db
  \phi_B(x_B,b_B)\phi^0 \big\{\big[(1-x_B-z)(1-\eta^2)\phi^A_K \nonumber\\
&-&rx_3(1-\eta)(\phi^P_K-\phi^T_K) +r(x_B+z)\eta(\phi^P_K+\phi^T_K)
  -2r\eta\phi^P_K\big] E_{2cd}(t_{2c})h_{2c}(x_B,z,x_3,b_B,b) -\big[(z-x_B \nonumber\\
&+&x_3(1-\eta))(1-\eta)\phi^A_K+r(x_B-z)\eta(\phi^P_K-\phi^T_K)
  -rx_3(1-\eta)(\phi^P_K+\phi^T_K) \big]E_{2cd}(t_{2d})h_{2d}(x_B,z,x_3,b_B,b)\big\}\;,
  \nonumber\\
  \\
M^{LR}_{TK} &=& 32\pi C_F m^4_B\sqrt{\eta}/\sqrt{2N_c} \int dx_B dz dx_3\int b_B db_B b db
  \phi_B(x_B,b_B) \big\{\big[(1-x_B-z)(1-\eta)(\phi^s+\phi^t)\phi^A_K \nonumber\\
&+& r(1-x_B-z)(\phi^s+\phi^t)(\phi^P_K-\phi^T_K) +r(x_3(1-\eta)+\eta)(\phi^s-\phi^t)
   (\phi^P_K+\phi^T_K) \big]E_{2cd}(t_{2c})h_{2c}(x_B,z,x_3,b_B,b)\nonumber\\
&-&\big[(z-x_B)(1-\eta)(\phi^s-\phi^t)\phi^A_K+r(z-x_B)(\phi^s-\phi^t)(\phi^P_K-\phi^T_K)
   +rx_3(1-\eta)(\phi^s+\phi^t)(\phi^P_K+\phi^T_K)\big]E_{2cd}(t_{2d}) \nonumber\\
&\times& h_{2d}(x_B,z,x_3,b_B,b)\big\}\;,\\
M^{SP}_{TK} &=& 32\pi C_F m^4_B/\sqrt{2N_c} \int dx_B dz dx_3\int b_B db_B b db \phi_B(x_B,b_B)\phi^0
   \big\{\big[(1+\eta-x_B-z+x_3(1-\eta))(1-\eta)\phi^A_K \nonumber\\
&+&r\eta(x_B+z)(\phi^P_K-\phi^T_K) -rx_3(1-\eta)(\phi^P_K+\phi^T_K)
   -2r\eta\phi^P_K\big] E_{2cd}(t_{2c})h_{2c}(x_B,z,x_3,b_B,b) \nonumber\\
&-&\left[(z-x_B)(1-\eta^2)\phi^A_K-rx_3(1-\eta)(\phi^P_K-\phi^T_K)
   +r\eta(x_B-z)(\phi^P_K+\phi^T_K)\right] E_{2cd}(t_{2d})h_{2d}(x_B,z,x_3,b_B,b)\big\}\;.
   \nonumber\\
\end{eqnarray}
The amplitudes from Fig.~1(d) are written as
\begin{eqnarray}
F^{LL}_{AK} &=& 8\pi C_F m^4_B f_B \int dz dx_3\int b db b_3 db_3\nonumber\\
&\times& \big\{\left[(x_3(1-\eta)-1)(1-\eta)\phi^A_K\phi^0 +2r\sqrt\eta(x_3(1-\eta)
    (\phi^P_K-\phi^T_K)-2\phi^P_K)\phi^s\right]E_{3ab}(t_{3a})h_{3a}(z,x_3,b,b_3) \nonumber\\
&+& \left[z(1-\eta)\phi^A_K\phi^0+2r\sqrt\eta\phi^P_K((1-\eta)(\phi^s-\phi^t)
    +z(\phi^s+\phi^t))\right]  E_{3ab}(t_{3b})h_{3b}(z,x_3,b,b_3) \big\}\;,\\
F^{SP}_{AK} &=& 16\pi C_F m^4_B f_B  \int dz dx_3\int b db b_3 db_3\nonumber\\
&\times& \big\{\left[2\sqrt\eta(1-\eta)\phi^A_K\phi^s +r(1-x_3)(\phi^P_K+\phi^T_K)\phi^0
    +r\eta((1+x_3)\phi^P_K-(1-x_3)\phi^T_K)\phi^0\right] E_{3ab}(t_{3a})h_{3a}(z,x_3,b,b_3)\nonumber\\
&+& \left[2r(1-\eta(1-z))\phi^P_K\phi^0+z\sqrt\eta((1-\eta)\phi^A_K(\phi^s-\phi^t) \right]
    E_{3ab}(t_{3b})h_{3b}(z,x_3,b,b_3)\big\}\;,\\
M^{LL}_{AK} &=& 32\pi C_F m^4_B/\sqrt{2N_c} \int dx_B dz dx_3\int b_B db_B b_3 db_3
  \phi_B(x_B,b_B) \big\{\big[(\eta-1)(-\eta+(1+\eta)(x_B+z))\phi^A_K\phi^0\nonumber\\
&+&r\sqrt\eta(x_3(1-\eta)+\eta)(\phi^P_K+\phi^T_K)(\phi^s-\phi^t)
  +r\sqrt\eta(1-x_B-z)(\phi^P_K-\phi^T_K)(\phi^s+\phi^t)-4r\sqrt\eta\phi^P_K\phi^s\big]
  E_{3cd}(t_{3c})\nonumber\\
&\times& h_{3c}(x_B,z,x_3,b_B,b_3)+\big[(1-\eta)((1-x_3)(1-\eta)-\eta(x_B-z))\phi^A_K\phi^0
  -r\sqrt\eta(x_B-z)(\phi^P_K+\phi^T_K)(\phi^s-\phi^t)\nonumber\\
&+&r\sqrt\eta(1-\eta)(1-x_3)(\phi^P_K-\phi^T_K)(\phi^s+\phi^t) \big]
   E_{3cd}(t_{3d})h_{3d}(x_B,z,x_3,b_B,b_3) \big\},\\
M^{LR}_{AK} &=&  32\pi C_F m^4_B/\sqrt{2N_c} \int dx_B dz dx_3\int b_B db_B b_3 db_3
  \phi_B(x_B,b_B)\big\{\big[\sqrt\eta(1-\eta)(2-x_B-z)\phi^A_K(\phi^s+\phi^t)-r(1+x_3)\nonumber\\
&\times&(\phi^P_K-\phi^T_K)\phi^0 -r\eta[(1-x_B-z)(\phi^P_K+\phi^T_K)
  -x_3(\phi^P_K-\phi^T_K)+2\phi^P_K]\phi^0\big]E_{3cd}(t_{3c})h_{3c}(x_B,z,x_3,b_B,b_3)\nonumber\\
&-&\big[r(1-\eta)(x_3- 1)(\phi^P_K-\phi^T_K)\phi^0
    +\sqrt\eta(x_B-z)[r\sqrt\eta(\phi^P_K+\phi^T_K)\phi^0+(1-\eta)\phi^A_K(\phi^s+\phi^t)] \big]
    E_{3cd}(t_{3d})\nonumber\\
&\times& h_{3d}(x_B,z,x_3,b_B,b_3)\big\}\;.
\end{eqnarray}
The hard functions $h_{i\alpha}$, the hard scales $t_{i\alpha}$, and the evolution factors
$E_{iab}$ and $E_{icd}$, with $i=1,2,3,4$ and $\alpha=a,b,c,d$, have their explicit expressions in the
Appendix of Ref.~\cite{prd89-074031}. 
Since the Legendre polynomial
$P_1(2\zeta-1)$ in the $P$-wave two-pion distribution amplitudes appears as an overall factor
in decay amplitudes, the integration over $\zeta$ can be performed trivially, yielding
a factor $\int_0^1 d\zeta(2\zeta-1)^2=1/3$ to branching ratios.


\end{document}